\documentclass[10pt, onecolumn]{article}
\usepackage[breaklinks=true]{hyperref}
\usepackage{amsmath}
\usepackage{authblk}
\usepackage{float}
\usepackage[margin=1in]{geometry}
\usepackage{graphicx}
\usepackage{tikz}
\usepackage[utf8]{inputenc}
\usepackage[numbers]{natbib}
\usepackage{tabularx}
\usepackage{xspace}
\usepackage{textcomp}
\newcommand{\ab}[1]{\textlangle#1\textrangle\xspace}

\newcolumntype{n}{X}
\newcolumntype{s}{>{\hsize=.35\hsize}X}

\restylefloat{table}

\title{Google COVID-19 Community Mobility Reports: Anonymization Process Description (version 1.1)}
\author{Ahmet Aktay}
\author{Shailesh Bavadekar}
\author{Gwen Cossoul}
\author{John Davis}
\author{Damien Desfontaines}
\author{Alex Fabrikant}
\author{Evgeniy Gabrilovich}
\author{Krishna Gadepalli}
\author{Bryant Gipson}
\author{Miguel Guevara}
\author{Chaitanya Kamath}
\author{Mansi Kansal}
\author{Ali Lange}
\author{Chinmoy Mandayam}
\author{Andrew Oplinger}
\author{Christopher Pluntke}
\author{Thomas Roessler}
\author{Arran Schlosberg}
\author{Tomer Shekel}
\author{Swapnil Vispute}
\author{Mia Vu}
\author{Gregory Wellenius}
\author{Brian Williams}
\author{Royce J Wilson}
\affil{community-mobility@google.com}

\begin{document}
\maketitle
\begin{abstract}
This document describes the aggregation and anonymization process applied to the initial version of Google COVID-19 Community Mobility Reports (published at http://google.com/covid19/mobility on April 2, 2020), a publicly available resource intended to help public health authorities understand what has changed in response to work-from-home, shelter-in-place, and other recommended policies aimed at flattening the curve of the COVID-19 pandemic. Our anonymization process is designed to ensure that no personal data, including an individual's location, movement, or contacts, can be derived from the resulting metrics. 

The high-level description of the procedure is as follows: we first generate a set of anonymized metrics from the data of Google users who opted in to Location History. Then, we compute percentage changes of these metrics from a baseline based on the historical part of the anonymized metrics. We then discard a subset which does not meet our bar for statistical reliability, and release the rest publicly in a format that compares the result to the private baseline.
\end{abstract}

\vfill

COVID-19 Community Mobility Reports provide insights into changes in mobility patterns. These reports use anonymized, aggregated data to chart movement trends over time by geography, as well as by place categories, showing trends over several weeks. This works in a similar way to existing Google products and features. For example, Google Maps uses aggregated, anonymized data to show how busy certain types of places are, including when a local business tends to be the most crowded. Public health officials have suggested this same type of aggregated, anonymized data could also be helpful as they make critical decisions to combat COVID-19. 

The COVID-19 Community Mobility Reports provide insights into what has changed in response to work-from-home, stay-at-home, and other recommended policies aimed at flattening the curve of the COVID-19 pandemic. They analyze trends in visits made to high-level categories of places, including workplaces, retail and recreational venues, groceries and pharmacies, parks, transit centers, and places of residence. Each version of the report will show trends over several weeks, with the most recent data representing 48 hours prior.

As explained in greater technical detail below, the anonymization process for these reports includes differential privacy~\cite{DP}, which is well-suited to produce analytics in contexts where the categories of data are known in advance. Our rigorous approach intentionally adds random noise to metrics in a way that maintains both users’ privacy and the overall accuracy of the aggregated data. 

\begin{figure}

\tikzset{every picture/.style={line width=0.75pt}} 

\begin{tikzpicture}[x=0.75pt,y=0.75pt,yscale=-1,xscale=1]

\draw   (20,30) rectangle (60,240) node[pos=.5,align=center,font=\tiny] {Original\\data} ;
\draw   (110,150) rectangle (200,190) node[pos=.5,align=center,font=\tiny] {Per-user contribution\\for non-baseline days} ;
\draw   (110,60) rectangle (200,100) node[pos=.5,align=center,font=\tiny] {Per-user contribution\\for baseline days} ;
\draw    (60,80) -- (107,80) ;
\draw [shift={(110,80)}, rotate = 180] [fill={rgb, 255:red, 0; green, 0; blue, 0 }  ][line width=0.08]  [draw opacity=0] (8.93,-4.29) -- (0,0) -- (8.93,4.29) -- cycle    ;
\draw    (60,170) -- (107,170) ;
\draw [shift={(110,170)}, rotate = 180] [fill={rgb, 255:red, 0; green, 0; blue, 0 }  ][line width=0.08]  [draw opacity=0] (8.93,-4.29) -- (0,0) -- (8.93,4.29) -- cycle    ;
\draw   (250,150) rectangle (340,190) node[pos=.5,align=center,font=\tiny] {Daily metrics\\for non-baseline days} ;
\draw   (250,60) rectangle (340,100) node[pos=.5,align=center,font=\tiny] {Daily metrics\\for baseline days} ;
\draw    (200,80) -- (247,80) ;
\draw [shift={(250,80)}, rotate = 180] [fill={rgb, 255:red, 0; green, 0; blue, 0 }  ][line width=0.08]  [draw opacity=0] (8.93,-4.29) -- (0,0) -- (8.93,4.29) -- cycle    ;
\draw    (200,170) -- (247,170) ;
\draw [shift={(250,170)}, rotate = 180] [fill={rgb, 255:red, 0; green, 0; blue, 0 }  ][line width=0.08]  [draw opacity=0] (8.93,-4.29) -- (0,0) -- (8.93,4.29) -- cycle    ;
\draw   (390,150) rectangle (480,190)node[pos=.5,align=center,font=\tiny] {Noisy metrics\\for non-baseline days} ;
\draw    (340,170) -- (387,170) ;
\draw [shift={(390,170)}, rotate = 180] [fill={rgb, 255:red, 0; green, 0; blue, 0 }  ][line width=0.08]  [draw opacity=0] (8.93,-4.29) -- (0,0) -- (8.93,4.29) -- cycle    ;
\draw   (390,60) rectangle (480,100) node[pos=.5,align=center,font=\tiny] {Noisy metrics\\for baseline days} ;
\draw    (340,80) -- (387,80) ;
\draw [shift={(390,80)}, rotate = 180] [fill={rgb, 255:red, 0; green, 0; blue, 0 }  ][line width=0.08]  [draw opacity=0] (8.93,-4.29) -- (0,0) -- (8.93,4.29) -- cycle    ;
\draw   (530,150) rectangle (620,190)node[pos=.5,align=center,font=\tiny] {Percentage changes\\between daily metrics\\and baseline metrics} ;
\draw    (480,170) -- (527,170) ;
\draw [shift={(530,170)}, rotate = 180] [fill={rgb, 255:red, 0; green, 0; blue, 0 }  ][line width=0.08]  [draw opacity=0] (8.93,-4.29) -- (0,0) -- (8.93,4.29) -- cycle    ;
\draw   (530,60) rectangle (620,100) node[pos=.5,align=center,font=\tiny] {Per-weekday median\\of baseline metrics} ;
\draw    (480,80) -- (527,80) ;
\draw [shift={(530,80)}, rotate = 180] [fill={rgb, 255:red, 0; green, 0; blue, 0 }  ][line width=0.08]  [draw opacity=0] (8.93,-4.29) -- (0,0) -- (8.93,4.29) -- cycle    ;
\draw   (530,240) rectangle (620,280) node[pos=.5,align=center,font=\tiny] {Published data} ;
\draw    (575,100) -- (575,147) ;
\draw [shift={(575,150)}, rotate = 270] [fill={rgb, 255:red, 0; green, 0; blue, 0 }  ][line width=0.08]  [draw opacity=0] (8.93,-4.29) -- (0,0) -- (8.93,4.29) -- cycle    ;
\draw    (575,190) -- (575,237) ;
\draw [shift={(575,240)}, rotate = 270] [fill={rgb, 255:red, 0; green, 0; blue, 0 }  ][line width=0.08]  [draw opacity=0] (8.93,-4.29) -- (0,0) -- (8.93,4.29) -- cycle    ;
\draw  [dash pattern={on 0.84pt off 2.51pt}]  (365,30) -- (365,240) ;

\draw (85,64.5) node  [font=\tiny] [align=left] {\begin{minipage}[lt]{40pt}\setlength\topsep{0pt}
\begin{center}
extract\\and bound
\end{center}

\end{minipage}};
\draw (85,154.5) node  [font=\tiny] [align=left] {\begin{minipage}[lt]{40pt}\setlength\topsep{0pt}
\begin{center}
extract\\and bound
\end{center}

\end{minipage}};
\draw (225,64.5) node  [font=\tiny] [align=left] {\begin{minipage}[lt]{40pt}\setlength\topsep{0pt}
\begin{center}
aggregate\\across users
\end{center}

\end{minipage}};
\draw (225,154.5) node  [font=\tiny] [align=left] {\begin{minipage}[lt]{40pt}\setlength\topsep{0pt}
\begin{center}
aggregate\\across users
\end{center}

\end{minipage}};
\draw (363.5,69.5) node  [font=\tiny] [align=left] {\begin{minipage}[lt]{40pt}\setlength\topsep{0pt}
\begin{center}
add noise
\end{center}

\end{minipage}};
\draw (363.5,159.5) node  [font=\tiny] [align=left] {\begin{minipage}[lt]{40pt}\setlength\topsep{0pt}
\begin{center}
add noise
\end{center}

\end{minipage}};
\draw (505,59.5) node  [font=\tiny] [align=left] {\begin{minipage}[lt]{40pt}\setlength\topsep{0pt}
\begin{center}
compute\\median per\\weekday
\end{center}
\end{minipage}};
\draw (402,230.5) node  [font=\tiny] [align=left] {\begin{minipage}[lt]{60pt}\setlength\topsep{0pt}
\begin{center}
privacy boundary
\end{center}
\end{minipage}};
\draw (529.5,135.5) node  [font=\tiny] [align=left] {\begin{minipage}[lt]{50pt}\setlength\topsep{0pt}
\begin{center}
calculate ratio
\end{center}
\end{minipage}};
\draw (553.5,211) node  [font=\tiny] [align=left] {\begin{minipage}[lt]{40pt}\setlength\topsep{0pt}
\begin{center}
filter\\unreliable\\metrics
\end{center}
\end{minipage}};
\end{tikzpicture}
\caption{System diagram of the metrics computation and anonymization process}\label{figure}

\end{figure}
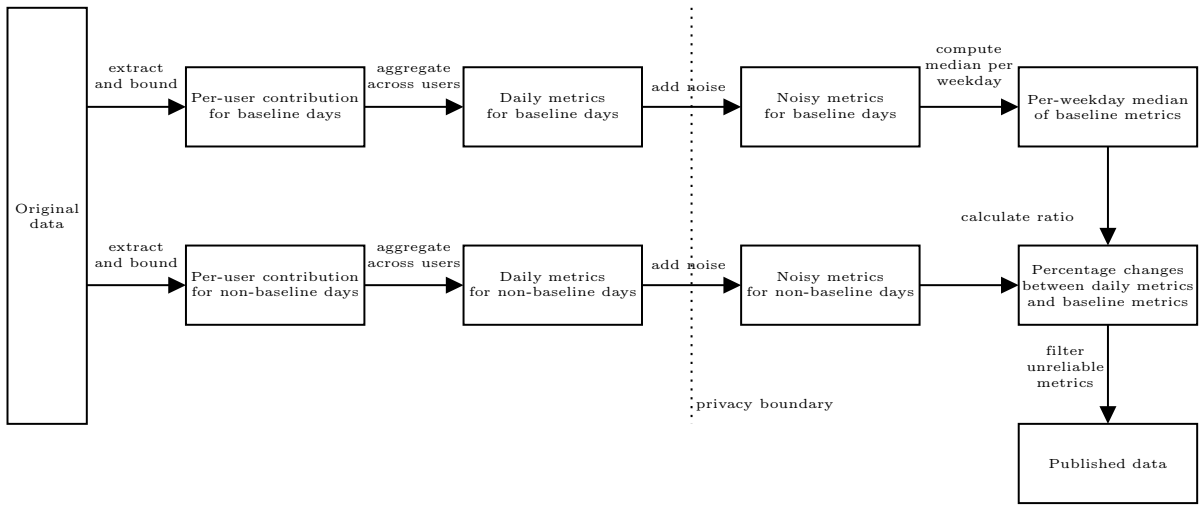

This paper is structured as follows: we introduce our method to produce anonymized metrics with differential privacy. We then explain how we post-process the anonymized metrics to generate the reports. Figure~\ref{figure} summarizes the anonymization process.

\section{Definitions}

\paragraph{Location History users}
The metrics in these reports are based on the data of Google users who have opted in to  \href{https://support.google.com/accounts/answer/3118687?hl=en}{Location History}~\cite{LH}, (``LH users''), a feature which is off by default.

\paragraph{Differential Privacy~\cite{DaMc06}}
Let $\varepsilon$ be a positive real number and A be a randomized algorithm that computes a metric. In the context of this report, A is considered $\varepsilon$-differentially private if for all input datasets $D_1$ and $D_2$ such that $D_2$ can be obtained from $D_1$ by adding or removing a single user's data in a single day, and for all subsets of $S \in \text{im}{A}$:
\begin{equation}
  Pr[A(D_1) \in S] \leq e^\varepsilon \cdot Pr[A(D_2) \in S].
  \nonumber
\end{equation}
\paragraph{Granularity levels}
The metrics are aggregated per day and per geographic area. There are three levels of geographic areas; in this paper, we call these granularity levels.
\begin{itemize}
  \item Granularity level 0 corresponds to metrics aggregated by country / region.
  \item Granularity level 1 corresponds to metrics aggregated by top-level geopolitical subdivisions (e.g. US states).
  \item Granularity level 2 corresponds to metrics aggregated by higher-resolution granularity (e.g. U.S. Counties).
\end{itemize}
Granularity levels 1 and 2 are defined differently in different countries, to account for knowledge of local public-health needs. Note that in general, the geographic area represented gets smaller as the granularity level increases. No metrics are published for geographic regions smaller than 3km$^2$.

\section{Generating anonymized metrics}
We are releasing aggregated, anonymized data that is designed to ensure that no personal data, including an individual’s location, movement, or contacts, can be derived from the resulting metrics. To that end, we anonymize the statistics with differential privacy. We query the underlying data using our open-source differential privacy library~\cite{dplibrary}, which adds Laplace noise~\cite{laplace} to protect each metric with differential privacy.

\subsection{Daily visits in public places}
We count the number of unique LH users who visited a public place of a given category in a given day at each granularity level. There are seven different categories derived from the data: retail, recreation, eateries (reported as part of “Retail \& recreation”); groceries, pharmacies; transit; and parks. We add Laplace noise to each count according to the following table.

\begin{table}[H]
  \small
  \centering
  \begin{tabularx}{0.70\linewidth}{c | c | c}
    Granularity level & Scale of Laplace noise & Corresponding $\varepsilon$ parameter
  \\
    \hline
    0 & $1/0.11\approx9.09 \quad (\sigma\approx12.86)$ & 0.11 \\
    1 & $1/0.11\approx9.09 \quad (\sigma\approx12.86)$ & 0.11   \\
    2 & $1/0.22 \approx4.55 \quad (\sigma \approx 6.43)$ & 0.22 \\
  \end{tabularx}
  \caption{\label{tab:publicvisits} Noise parameters used for the daily visits in public places metrics}
\end{table}
For each location (at all geographic levels), each LH user can contribute at most once to each category. We also bound the contribution of each LH user to 4 \ab{category,location} pairs per day and per geographic level, using a process similar to the one described in this paper~\cite{dpsql}: if an LH user contributes to more than 4 pairs in a given day and given geographic level, we randomly select 4 of them, and discard the others.

For example, suppose that on the same day, an LH user goes to public places in all 7 categories in two distinct neighboring countries. This makes a total of 14 \ab{category,location} pairs at country level.  We would randomly discard 10 of these pairs when computing country-level statistics.

This process does not significantly affect data accuracy: in the US, at county level, $99\%$ of LH users contribute 3 or fewer \ab{category,place} pairs per day on average.
Thus, each daily place visit is protected by differential privacy with $\varepsilon=0.44$. These multiple metrics apply to the same dataset, so standard composition results apply, and the total daily contribution of each user is protected by differential privacy with a maximum of $\varepsilon=1.76$.

\subsection{Residential}
For the purposes of this analysis, we use signals like relative frequency, time and duration of visits to calculate metrics related to places of residence. We calculate an average amount of time spent at places of residence for LH users in hours. This computation is performed for each day and geographic area, using the same algorithm as the differentially private mean mechanism from our open-source library~\cite{dpboundedmean}. This mechanism works as follows:
\begin{itemize}
    \item We compute the amount of time spent at place of residence in a given day and geographic area in hours by summing up the individual values per user offset by 12, so all individual values fall into the range $[-12; 12]$. We then add Laplace noise to this sum; the scale of the noise is indicated in the table below. We denote the real sum $s$, and noisy sum $s_n$.
    \item We compute the count of unique users who spent any time at residences in a given day and geographic area. We then add Laplace noise to this count; the scale of the noise is indicated in the table below. We refer to the real count $c$, and the noisy count $c_n$.
    \item Finally, we compute the ratio $s_n/c_n$ for each day and each geographic area, add 12 as offset, and clamp it to the range $[0, 24]$ hours/day.
\end{itemize}
For example, at county-level, $s_n$ is obtained by first sampling a random number from a Laplace distribution of scale 109.1, and then adding that number to $s$. In the table below, we also indicate the standard deviation $\sigma$ of the noise added to each value.
\begin{table}[H]
  \small
  \centering
  \begin{tabularx}{0.9\linewidth}{c | c | c | c}
    Granularity level & Scale of Laplace noise: & Scale of Laplace noise: & Corresponding \\
    & sum (total hours/day) &  count (number of users) & $\varepsilon$ parameter
    \\
    \hline
    0 & $12/0.055 \approx 218.2 \quad (\sigma \approx 308.6)$ & $1/0.055\approx18.2 \quad (\sigma\approx25.71)$ & 0.11   \\
    1 & $12/0.055\approx218.2 \quad (\sigma\approx308.6)$ & $1/0.055\approx18.2 \quad (\sigma\approx25.71)$ & 0.11   \\
    2 & $12/0.110\approx109.1 \quad (\sigma\approx154.3)$ & $1/0.110\approx9.09 \quad (\sigma\approx12.86)$ & 0.22  
  \end{tabularx}
  \caption{\label{tab:residential} Noise parameters used for the residential metrics}
\end{table}
Each user can contribute to at most one region per granularity level, which protects these metrics by differential privacy with $\varepsilon=0.44$ total budget across all granularities. A description of the differentially private mean mechanism implemented and a proof of its privacy guarantees is described in~\cite{dptheorypractice} (Algorithm 2.4).

\subsection{Workplaces}
For the purposes of this analysis, we use signals like relative frequency, time and duration of visits to calculate metrics related to places of residence and places of work of LH users. We calculate how many LH users spent more than 1 hour at their places of work. This computation is performed for each day and geographic area. Then, we add Laplace noise to each count according to the following table. 
\begin{table}[H]
  \small
  \centering
  \begin{tabularx}{0.7\linewidth}{c | c | c}
    Geographic level & Scale of Laplace noise  & Corresponding $\varepsilon$ parameter
    \\
    \hline
    0 & $1/0.11 \approx 9.09 \quad (\sigma \approx 12.86)$ & 0.11 \\
    1 & $1/0.11\approx 9.09 \quad (\sigma \approx 12.86)$ & 0.11 \\
    2 & $1/0.22 \approx 4.55 \quad (\sigma \approx 6.43)$ & 0.22 \\
  \end{tabularx}
  \caption{\label{tab:workplaces} Noise parameters used for the work places metrics}
\end{table}
The count is aggregated by places of residence of LH users. Since each user can contribute to at most one geographic area per granularity level, these metrics are protected by differential privacy with $\varepsilon=0.44$.

\section{Generating the report from the anonymized metrics}
The metrics described above are generated for each day, starting on 2020-01-01. They are then used to generate the percentage changes relative to day of the week published in the reports. All operations described below use only the output of the differentially private mechanisms described in the previous section; so they do not consume any privacy budget.

\paragraph{Additional privacy protections}
We discard all metrics for which the geographic region is smaller than 3km$^2$, or for which the differentially private count of contributing users (after noise addition) is smaller than 100. Geographic regions smaller than 3km$^2$ may be merged such that the union of their area is above the 3km$^2$ threshold. This merging does not occur across country boundaries, except for the Vatican City and Italy.

\subsection{Computing percentage changes from a baseline}

For each individual metric generated using the mechanisms described above, we compute the ratio between the metric for a given day D and the same metric computed for the baseline period. The reference baseline is defined in the following way.
\begin{itemize}
    \item We consider the 5-week range from 2020-01-03 through 2020-02-06. This ranged is fixed.
    \item Within this 5-week range, we consider the 5 days with the same day of week as $d$. For example, if $d$ is 2020-03-20, $d$ is a Friday, so we consider the 5 Fridays in this 5-week range (Jan 3 to Jan 31, inclusive).
    \item We compute the median of the differentially private metrics for these 5 baseline days.
    \item This median metric is the baseline metric for $d$.
\end{itemize}
We then compute and publish the ratio between the metric for $d$ and the baseline metric, as a percentage.

\subsection{Removing unreliable metrics}
In some regions, the noise added to obtain differential privacy can reduce the confidence that we are capturing a meaningful change, typically when there is not a lot of data for the metric. When, because of this uncertainty, the percentage change for one of these metrics has a $5\%$ chance (or higher) of being wrong by more than $\pm10$ absolute percentage points, we do not publish it and instead include an asterisk denoting that there is not enough data available to present privacy-safe information. More precisely:
\begin{itemize}
\item Before releasing a ratio metric/baseline, we compute $97.5\%$ confidence intervals for the metric and its baseline. Let us denote [$m_{\min}$, $m_{\max}$] and [$b_{\min}$, $b_{\max}$] these respective confidence intervals.
\item We compute the ratios $m_{\min}/b_{\max}$ and  $m_{\max}/b_{\min}$.
\item If one of these ratios differs from the differentially private ratio by more than 10 absolute percentage points, we do not publish the corresponding percentage changes.
\end{itemize}
If the last condition is not satisfied, then the probability of being wrong by more than 10 absolute percentage points in each direction is lower than 2.5\%. By union bound, this means that there is at most a 5\% risk of being wrong by more than 10 absolute percentage points.
Note that the confidence intervals are based on an already differentially private value and on public data (the scale and shape of the noise), so no privacy budget is consumed by this operation.

\section{Note on $\delta$}

We are generating a fixed set of metrics (all possible combinations of geographic regions, days within the periods, and public place categories), and we are also adding noise to zero-valued metrics. As such, the process outlined above is $\varepsilon$-differentially private with $\delta=0$.

\section{Improving the accuracy of metrics over time}

We are continuously making improvements to the underlying computation of the metrics to improve their accuracy over time. These updates can introduce a shift in their values, which can skew comparisons over time compared to the baseline period if they are not accounted for, since those improvements are not applied to the baseline period (to avoid republishing the data). To roll out these changes with minimal impact on the overall privacy budget, we use \emph{scaling factors}. Rather than recomputing a metric directly, which would require a larger value of $\varepsilon$, we use the following process.
\begin{enumerate}
\item We define groups of metrics, on which the effect of the update is uniform. For example, for a specific metric and location, we could consider all days in a given period, or group days by weekday within this period (e.g. Parks metric across all Tuesdays in June in the US, or Workplaces metric across week-ends in August in each region of granularity level 1).
\item For each group, we sum the noisy metrics already generated with the previous computation logic. Call this sum $s_g$.
\item For each group, we sum the metrics recomputed for the same given period with the new computation logic, then we add noise to it with a smaller privacy budget (typically, 10\%), proportional to the budget we use for the corresponding region granularity. Call this noisy sum $s_n$.
\item For dates following the given period, we multiply the metrics generated with the new logic by the scaling factor $s_n/s_g$ (if we are scaling the baseline) or by its inverse $s_g/s_n$ (if we are scaling the daily counts).
\end{enumerate}

Grouping multiple metrics together to compute this scaling factor is the key insight that allows us to use a much smaller privacy budget for step 3, and reuse already generated metrics in step 2.

\bibliographystyle{unsrt}
\bibliography{references}
\end{document}